\documentclass[12pt]{article}

\usepackage{amsmath}
\usepackage{amssymb}
\usepackage[dvips]{graphicx}
\usepackage[dvips]{psfrag}
\usepackage{cite}
\usepackage{subfigure}

\makeatletter

  \@addtoreset{equation}{section}
 \makeatother

\newcommand {\n}{\nonumber \\}

\begin{document}
\setlength{\oddsidemargin}{0cm}
\setlength{\baselineskip}{7mm}

\begin{titlepage}
\begin{normalsize}
\begin{flushright}
\begin{tabular}{l}
February 2011
\end{tabular}
\end{flushright}
  \end{normalsize}

~~\\

\vspace*{0cm}
    \begin{Large}
       \begin{center}
         {Zariski Quantization as Second Quantization}
       \end{center}
    \end{Large}
\vspace{1cm}

\begin{center}
           Matsuo S{\sc ato}\footnote
           {
e-mail address : msato@cc.hirosaki-u.ac.jp}\\
      \vspace{1cm}
       
         {\it Department of Natural Science, Faculty of Education, Hirosaki University\\ 
 Bunkyo-cho 1, Hirosaki, Aomori 036-8560, Japan}

\end{center}

\hspace{5cm}

\begin{abstract}
\noindent
The Zariski quantization is one of the strong candidates for a quantization of the Nambu-Poisson bracket. In this paper, we apply the Zariski quantization for first quantized field theories, such as superstring and supermembrane theories,  and clarify physical meaning of the Zariski quantization. The first quantized field theories need not to possess the Nambu-Poisson structure. First, we construct a natural metric for the spaces on which Zariski product acts in order to apply the Zariski quantization for field theories. This metric is invariant under a gauge transformation generated by the Zariski quantized Nambu-Poisson bracket. Second, we perform the Zariski quantization of superstring and supermembrane theories as examples. We find flat directions, which indicate that the Zariski quantized theories describe many-body systems. We also find that pair creations and annihilations occur among the many bodies introduced by the Zariski quantization, by studying a simple model. These facts imply that the Zariski quantization is a second quantization. Moreover, the Zariski quantization preserves supersymmetries of the first quantized field theories. Thus, we can obtain second quantized theories of superstring and supermembranes by performing the Zariski quantization of the superstring and supermembrane theories. 
\end{abstract}

\vfill
\end{titlepage}
\vfil\eject

\setcounter{footnote}{0}

\section{Introduction}
\setcounter{equation}{0}
Quantization of the Nambu-Poisson bracket has been a long-standing problem since Y. Nambu proposed in 1973 \cite{Nambu}. The Zariski quantization proposed in 1996 by G. Dito, M. Flato, D. Sternheimer, and L. Takhtajan, is one of the strong candidates for a quantization of the Nambu-Poisson bracket \cite{DitoFlatoSternheimerTakhtajan}. The Zariski quantization consists of two steps. First, instead of a direct quantization of the original Nambu-Poisson bracket, they deform spaces on which the Nambu-Poisson bracket acts in the classical level, and define a new Nambu-Poisson bracket that has the same Nambu-Poisson structure as the original one. The definition of the Nambu-Poisson structure is to satisfy  the Leibniz rule, the skew-symmetry, and  the Fundamental Identity, which is a generalization of the Jacobi identity. Second, they define a deformation quantization of the new Nambu-Poisson bracket. Because the Zariski quantized Nambu-Poisson bracket reduces not to the original Nambu-Poisson bracket but to the new Nambu-Poisson bracket in the classical limit, the Zariski quantization is regarded as different with usual quantizations \cite{Yoneya}. For this reason, applications for Physics have developed little for a long time.   

In this paper, we study a relation between the original and the new Nambu-Poisson brackets and clarify physical meaning of the Zariski quantization. In subsection 2.1, we summarize definitions and mathematical properties of the classical Zariski product and spaces $\mathcal{M}$ on which it acts. We also define the new Nambu-Poisson bracket on $\mathcal{M}$ and show that it possesses the same Nambu-Poisson structure as the original one, whereas G. Dito et al. defined it on $\mathcal{A}_0$ that are small subspaces of $\mathcal{M}$. It is necessary for later discussions to define the Zariski quantization not on $\mathcal{A}_0$ but on $\mathcal{M}$. In subsection 2.2, we construct a natural metric for $\mathcal{M}$ in order to apply the Zariski quantization for field theories. We show that the metric is invariant under a gauge transformation generated by the new Nambu-Poisson bracket, although 3-algebras equipped with invariant metrics are restricted in general. In subsection 2.3, we deform superstring and supermembrane theories by the classical Zariski product, the spaces $\mathcal{M}$, and the metric. This deformation can be applied for any first quantized field theory, even if it does not possess the Nambu-Poisson structure. We show that the deformed actions reduce to the actions before the deformation if we restrict their fields to only one-body states. We also show that the deformed actions possess flat directions. These results imply that the deformation of first quantized theories by the classical Zariski product, the space $\mathcal{M}$, and the metric is a many-body deformation. As a corollary, we find that the new Nambu-Poisson bracket is the many-body deformation of the original Nambu-Poisson bracket.  

In subsection 3.1, we summarize definitions and mathematical properties of the deformation quantization in the Zariski quantization. By performing the deformation quantization of the classical Zariski product and $\mathcal{M}$, we obtain the quantum Zariski product and $\mathcal{M}_{\hbar}$. We can define the Zariski quantized Nambu-Poisson bracket by the deformation quantization of the many-body deformation of the original Nambu-Poisson bracket. In subsection 3.2, we construct a natural metric for $\mathcal{M}_{\hbar}$. This metric is invariant under a gauge transformation generated by the Zariski quantized Nambu-Poisson bracket. In subsection 3.3, we perform the Zariski quantization of first quantized field theories and study their general features by using a simple model. The Zariski quantization is applicable to any first quantized field theory, which is not necessary to possess the Nambu-Poisson structure. We define theories by path-integrals of Zariski quantized actions, which are $\hbar$ deformations of classical actions. As a result, we find that pair creations and annihilations occur among the many bodies that are introduced by the many-body deformation. Therefore, by performing the Zariski quantization, which consists of the many-body deformation and the deformation quantization, we obtain second quantized theories from first quantized field theories. The Zariski quantization preserves supersymmetries of first quantized  field theories, because the quantum Zariski product is Abelian, associative and distributive, and admits a commutative derivative satisfying the Leibniz rule. Therefore, by performing  the Zariski quantization of superstring and supermembrane theories, we can obtain second quantized theories of the superstrings and the supermembranes.  

\vspace{1cm}

\section{Classical Zariski Product and Many-Body Deformation}
\setcounter{equation}{0}
In this section, we find that actions representing single body systems become those representing many-body systems, by replacing product and spaces on which it acts, with the classical Zariski product and $\mathcal{M}$. As a corollary, we find that the new Nambu-Poisson bracket defined by using the classical Zariski product and $\mathcal{M}$, is a many-body deformation of the original Nambu-Poisson bracket.

\subsection{Definitions and Mathematical Properties}
In this subsection, we summarize definitions and mathematical properties of the classical Zariski product \cite{DitoFlatoSternheimerTakhtajan} and spaces $\mathcal{M}$ on which it acts. We also define a Nambu-Poisson bracket deformed by the classical Zariski product on $\mathcal{M}$, which will be interpreted as many-body spaces later, instead of the small subspaces $\mathcal{A}_0$ defined in \cite{DitoFlatoSternheimerTakhtajan}.

First, we define elements of linear spaces $\mathcal{M}$ by
\begin{equation}
\bold{X}=\sum_{u} Y_{u}(\sigma) Z_{u}, 
\end{equation}
where the basis $Z_{u}$ are labeled by polynomials $u=u(x_1,x_2)$ in the valuables $x_1$, $x_2$ with real coefficients. $Z_{u}$ satisfies $Z_{au}=aZ_{u}$ where $a$ is a real number. The coefficients $Y_{u}(\sigma)$ are functions over $p$-dimensional spaces. Summation is defined naturally as linear spaces. 

The classical Zariski product $\bullet$ is defined on $\mathcal{M}$ by 
\begin{eqnarray}
\bold{X} \bullet \bold{X}'&=& (\sum_{u} Y_{u}(\sigma) Z_{u}) \bullet (\sum_{v} Y'_{v}(\sigma) Z_{v}) \n&=& \sum_{u, v} Y_{u}(\sigma) Y'_{v}(\sigma) Z_{uv}. 
\end{eqnarray}
The classical Zariski product is Abelian, distributive and associative as follows:
\begin{equation}
\bold{X} \bullet \bold{X}'= \sum_{u, v} Y_{u}(\sigma) Y'_{v}(\sigma) Z_{uv}
=\sum_{v, u}  Y'_{v}(\sigma) Y_{u}(\sigma) Z_{vu}
=\bold{X}' \bullet \bold{X}, \label{classicalabelian}
\end{equation}
\begin{eqnarray}
\bold{X} \bullet (\bold{X}'+\bold{X}'')&=& \sum_{u} Y_{u}(\sigma) Z_{u} \bullet
(\sum_{v} Y'_{v}(\sigma) Z_{v}+ \sum_{w} Y''_{w}(\sigma) Z_{w}) \n
&=&\sum_{u, v} Y_{u}(\sigma) Y'_{v}(\sigma) Z_{uv}
+\sum_{u, w} Y_{u}(\sigma) Y''_{w}(\sigma) Z_{uw} = \bold{X} \bullet \bold{X}'+\bold{X} \bullet \bold{X}'',
\end{eqnarray}
and
\begin{eqnarray}
(\bold{X} \bullet \bold{X}') \bullet \bold{X}''
&=&(\sum_{u, v} Y_{u}(\sigma) Y'_{v}(\sigma) Z_{uv})
\bullet(\sum_{w} Y''_{w}(\sigma) Z_{w}) \n
&=& \sum_{u, v, w} Y_{u}(\sigma) Y'_{v}(\sigma) Y''_{w}(\sigma) Z_{uvw}
=\bold{X} \bullet (\bold{X}' \bullet \bold{X}'').
\end{eqnarray}

We define derivatives on $\mathcal{M}$ by derivatives with respect to $\sigma^i$ ($i=1,2, \cdots, p$). These derivatives are commutative:
\begin{equation}
\frac{\partial}{\partial \sigma^i}\frac{\partial}{\partial \sigma^j}\bold{X}
=\sum_{u} (\frac{\partial}{\partial \sigma^i}\frac{\partial}{\partial \sigma^j}Y_{u}(\sigma)) Z_{u}
=\sum_{u} (\frac{\partial}{\partial \sigma^j}\frac{\partial}{\partial \sigma^i}Y_{u}(\sigma)) Z_{u}
= \frac{\partial}{\partial \sigma^j}\frac{\partial}{\partial \sigma^i}\bold{X},
\end{equation}
and the derivatives of the classical Zariski product satisfy the Leibniz rule:
\begin{eqnarray}
\frac{\partial}{\partial \sigma^i}(\bold{X} \bullet \bold{X}')
&=&\sum_{u, v} \frac{\partial}{\partial \sigma^i}(Y_{u}(\sigma) Y'_{v}(\sigma)) Z_{uv} \n
&=&\sum_{u, v} (\frac{\partial}{\partial \sigma^i}Y_{u}(\sigma) Y'_{v}(\sigma) +Y_{u}(\sigma) \frac{\partial}{\partial \sigma^i}Y'_{v})Z_{uv} \n
&=& \sum_{u} \frac{\partial}{\partial \sigma^i}Y_{u}(\sigma) Z_{u} 
\bullet \sum_{v} Y'_{v}(\sigma) Z_{v} 
+\sum_{u} Y_{u}(\sigma) Z_{u} 
\bullet \sum_{v} \frac{\partial}{\partial \sigma^i}Y'_{v}(\sigma) Z_{v} \n
&=&\frac{\partial}{\partial \sigma^i}\bold{X} \bullet \bold{X}'+ \bold{X} \bullet \frac{\partial}{\partial \sigma^i}\bold{X}'. \label{classicalLeibnitz}
\end{eqnarray}
We define a Nambu-Poisson bracket deformed by the classical Zariski product and $\mathcal{M}$ by
\begin{eqnarray}
[\bold{X}, \bold{X}', \bold{X}'']_{\bullet}
&:=&\epsilon^{ijk}
\frac{\partial}{\partial\sigma^i} \bold{X} \bullet
\frac{\partial}{\partial\sigma^j} \bold{X}'\bullet
\frac{\partial}{\partial\sigma^k} \bold{X}'' \n
&=&
\sum_{u,v,w}
\epsilon^{ijk}
\frac{\partial}{\partial\sigma^i} Y_{u}(\sigma)
\frac{\partial}{\partial\sigma^j} Y'_{v}(\sigma)
\frac{\partial}{\partial\sigma^k} Y''_{w}(\sigma)
Z_{uvw},
\end{eqnarray}
where $i, j, k$ run from 1 to 3.
By definition, the bracket is skew-symmetric, that is, totally anti-symmetric in all the three entries.  By using (\ref{classicalabelian}) - (\ref{classicalLeibnitz}), one can easily show that it satisfies the Leibniz rule and the Fundamental Identity;
\begin{equation}
\{A, B,\{X,Y,Z\}\}=\{\{A,B,X\},Y,Z\}+\{X,\{A,B,Y\},Z\}+\{X,Y,\{A,B,Z\}\},
\end{equation}
for any $A, B, X, Y, Z \in \mathcal{M}$. Thus, the deformed Nambu-Poisson bracket possesses the same Nambu-Poisson structure as the original one.

\subsection{Metric}
In this subsection, we construct a natural metric for $\mathcal{M}$, in order to apply the classical Zariski product and $\mathcal{M}$ for field theories.

We define a metric for $X, X' \in \mathcal{M}$ by
\begin{eqnarray}
<\bold{X}, \bold{X}'>
&=&
<\bold{X} \bullet \bold{X}'> \n
&=&
\int d^p\sigma <<\bold{X} \bullet \bold{X}'>> \n
&=&
\sum_{u,v} \int d^p \sigma Y_{u}(\sigma) Y'_{v}(\sigma) <<Z_{uv}>>.
\end{eqnarray}
$<<Z_{w}>>$ is defined by
\begin{equation}
<<Z_{w}>>=a \mbox{ if } w=a z^2, \quad \mbox{otherwise} <<Z_{w}>>=0,
\end{equation}
where $a$ is a real number and $z$ is a normalized polynomial, whose monomial of the highest total degree has coefficient 1. 

This metric is invariant under a gauge transformation generated by the $p$-dimensional deformed Nambu-Poisson bracket,  
$[\bold{X}^1, \bold{X}^2, \cdots, \bold{X}^p]_{\bullet}
:=\epsilon^{i_1 i_2 \cdots i_p}
\frac{\partial}{\partial\sigma^{i_1}} \bold{X}^1 \bullet
\frac{\partial}{\partial\sigma^{i_2}} \bold{X}^2 \bullet \cdots \bullet
\frac{\partial}{\partial\sigma^{i_p}} \bold{X}^p$.
Here we show the $p=3$ case as an example, whereas the $p=2$ case (Poisson bracket) and the $p>3$ case are shown in a similar way. The condition of the invariance of the metric is given by   
\begin{equation} 
\delta <\bold{X}^1, \bold{X}^2>
=
<\delta \bold{X}^1, \bold{X}^2> + <\bold{X}^1, \delta \bold{X}^2> =0.
\end{equation}
This is equivalent to
\begin{equation}
<[\bold{X}^3, \bold{X}^4, \bold{X}^1]_{\bullet}, \bold{X}^2> + <\bold{X}^1, [\bold{X}^3, \bold{X}^4, \bold{X}^2]_{\bullet}>=0.
\end{equation}
The left hand side is
\begin{eqnarray}
&&<\sum_{u_3,u_4,u_1}
\epsilon^{ijk}
\frac{\partial}{\partial\sigma^i} Y^3_{u_3}(\sigma)
\frac{\partial}{\partial\sigma^j} Y^4_{u_4}(\sigma)
\frac{\partial}{\partial\sigma^k} Y^1_{u_1}(\sigma)
Z_{u_3 u_4 u_1}
\sum_{u_2} Y^2_{u_2}(\sigma) Z_{u_2}> \n
&&+
<\sum_{u_1} Y^1_{u_1} (\sigma) Z_{u_1},
\sum_{u_3,u_4,u_2}
\epsilon^{ijk}
\frac{\partial}{\partial\sigma^i} Y^3_{u_3}(\sigma)
\frac{\partial}{\partial\sigma^j} Y^4_{u_4}(\sigma)
\frac{\partial}{\partial\sigma^k} Y^2_{u_2}(\sigma)
Z_{u_3 u_4 u_2}> \n
&=&
\sum_{u_1, u_2, u_3, u_4}
\int d^3 \sigma (\epsilon^{ijk}
\frac{\partial}{\partial\sigma^i} Y^3_{u_3}(\sigma)
\frac{\partial}{\partial\sigma^j} Y^4_{u_4}(\sigma)
\frac{\partial}{\partial\sigma^k} Y^1_{u_1}(\sigma)
Y^2_{u_2}(\sigma) \n
&& \qquad \qquad \qquad+
Y^1_{u_1} (\sigma)
\epsilon^{ijk}
\frac{\partial}{\partial\sigma^i} Y^3_{u_3}(\sigma)
\frac{\partial}{\partial\sigma^j} Y^4_{u_4}(\sigma)
\frac{\partial}{\partial\sigma^k} Y^2_{u_2}(\sigma))
<<Z_{u_1 u_2 u_3 u_4}>> \n
&=&
\sum_{u_1, u_2, u_3, u_4}
\int d^3 \sigma (
\frac{\partial}{\partial\sigma^k}
(\epsilon^{ijk}
\frac{\partial}{\partial\sigma^i} Y^3_{u_3}(\sigma)
\frac{\partial}{\partial\sigma^j} Y^4_{u_4}(\sigma)
Y^1_{u_1}(\sigma)
Y^2_{u_2}(\sigma)) )
<<Z_{u_1 u_2 u_3 u_4}>> \n
&=&0.
\end{eqnarray}

\subsection{Flat Directions}
In this subsection, we deform first quantized field theories, superstring  and supermembrane theories as examples, by using the classical Zariski product and $\mathcal{M}$. We study ground states and find flat directions, which indicates the deformed theories describe many-body systems.

We can deform any first quantized action $S=\int d^p \sigma \mathcal{L}(X)$ into $S=< \mathcal{L}(\bold{X})_{\bullet}>$. The action need not to possess the Nambu-Poisson structure. If we restrict $\bold{X} = \sum_{u} Y(\sigma)_u Z_u$ into $\bold{X}=Y(\sigma)_v Z_v$, $S=< \mathcal{L}(\bold{X})_{\bullet}>$ reduces to $S=\int d^p \sigma \mathcal{L}(Y_v)$ because $<<Z_{v^2}>>$ is a non-zero constant. This implies that each $Y(\sigma)_v Z_v$ among $\sum_{u} Y(\sigma)_u Z_u$ should be a single body state. Irreducible polynomials can label single particle state whereas reducible polynomials can label bound states. 

In order to examine whether $S=< \mathcal{L}(\bold{X})_{\bullet}>$ represent many-body systems, we study ground states in deformed superstring and supermembrane theories as examples. The bosonic part of the superstring Hamiltonian in a light-cone gauge is given by 
\begin{equation}
H=\frac{l}{4\pi\alpha'p^+} \int_0^l d \sigma 
(\frac{2\pi\alpha' (p^+)^2}{l^2}(\partial_{\tau} X^i)^2 
+\frac{1}{2\pi \alpha'}(\partial_{\sigma} X^i)^2).
\end{equation}
After the deformation, we obtain
\begin{equation}
H_{\bullet}=\frac{l}{4\pi\alpha'p^+} 
< \frac{2\pi\alpha' (p^+)^2}{l^2} \partial_{\tau} \bold{X}^i \bullet \partial_{\tau} \bold{X}^i
+\frac{1}{2\pi \alpha'} \partial_{\sigma} \bold{X}^i \bullet \partial_{\sigma} \bold{X}^i >,
\end{equation}
where $\bold{X}^i = \sum_u Y^i_u (\tau, \sigma) Z_u$. 
$H_{\bullet}=0$ for 
\begin{equation}
\bold{X}^i = \sum_u Y^i_u Z_u,
\end{equation}
where $Y^i_u$ are constants. This means $\{Y^i_u\}$ represent flat directions. One can show that these flat directions are preserved after a deformation quantization that will be defined in the next section. Moreover, the deformed theory preserves the supersymmetry of the superstring theory, because the classical and quantum Zariski products are Abelian, distributive, and associative, and admit commutative derivatives satisfying the Leibniz rule. Thus, quantum corrections to the flat directions should be suppressed. Therefore, this theory possesses a continuous spectrum and describes a many-body system. The flat directions correspond to positions of many-body superstrings.

One can also apply the deformation for the supermembrane Hamiltonian in a light-cone gauge
 and obtains 
\begin{equation}
H_{\bullet}=\frac{\nu T}{4} < (\partial_{\tau} \bold{X}^i)^2_{\bullet}
+\frac{2}{\nu^2} \{ \bold{X}^i, \bold{X}^j \}^2_{\bullet}
-\frac{2}{\nu} \Theta^T \bullet \gamma_i \{X^i, \Theta \}_{\bullet}>,
\end{equation}
where $\bold{X}^i = \sum_u Y^i_u (\tau, \sigma) Z_u$ and 
$\Theta = \sum_u \theta_u (\tau, \sigma) Z_u$.
%
$H_{\bullet}=0$ for
\begin{equation}
\bold{X}^i = \sum_u Y^i_u Z_u, \qquad \Theta=0,
\end{equation}
where $Y^i_u$ are constants. 
Thus, $\{Y^i_u\}$ are flat directions, which correspond to positions of many-body membranes.  

From these general features, we conclude that the deformations of first quantized field theories by the classical Zariski product and $\mathcal{M}$ are many-body deformations. 

\section{Zariski Quantization as Second Quantization}
\setcounter{equation}{0}
In this section, we perform a deformation quantization of the classical Zariski product and $\mathcal{M}$, and obtain Zariski quantized actions from the deformed actions. We define quantum theories of the Zariski quantized actions by using a path-integral. We find that pair creations and annihilations occur among the many bodies that are introduced by the many-body deformation. This indicates that first quantized field theories become second quantized theories by the Zariski quantization.

\subsection{Definitions and Mathematical Properties}
In this subsection, we summarize definitions and mathematical properties of a deformation quantization in the Zariski quantization \cite{DitoFlatoSternheimerTakhtajan}.

The deformation quantization of $\mathcal{M}$ is defined by   
\begin{equation}
\bold{X}_{\alpha}=\sum_{r=0}^{\infty} \alpha^r \sum_{u_r} Y^r_{u_r}(\sigma) Z_{u_r} \in \mathcal{M}_{\hbar},\end{equation}
where $\alpha$ is a deformation parameter related to $\hbar$. We will determine the relation later.
The quantum Zariski product $\bullet_{\hbar}$ is defined by a deformation quantization of the classical Zariski product as
\begin{eqnarray}
\bold{X}_{\hbar} \bullet_{\hbar} \bold{X}_{\hbar}'&=& (\sum_{r=0}^{\infty} \alpha^r \sum_{u_r} Y^r_{u_r}(\sigma) Z_{u_r})\bullet_{\hbar} (\sum_{s=0}^{\infty} \alpha^s \sum_{v_s} Y'^s_{v_s}(\sigma) Z_{v_s}) \n
&=& (\sum_{u_0} Y^0_{u_0}(\sigma) Z_{u_0})
\bullet_{\hbar} 
(\sum_{v_0} Y'^0_{v_0}(\sigma) Z_{v_0}) \n
&=& \sum_{u_0, v_0} Y^0_{u_0}(\sigma) Y'^0_{v_0}(\sigma)  Z_{u_0} \bullet_{\hbar} Z_{v_0}. \label{quantumproduct1}
\end{eqnarray}
Any polynomial can be decomposed uniquely as $u=a u_1 u_2 \cdots u_M$, where $a$ is a real number and $u_i$ are irreducible normalized polynomials. $Z_u \bullet_{\hbar} Z_v$ is defined by 
\begin{equation}
Z_u \bullet_{\hbar} Z_v
=
ab \zeta((u_1 u_2 \cdots u_M) \times_{\hbar} (u_{M+1}u_{M+2} \cdots u_{N})), \label{quantumproduct2}
\end{equation}
where $v=b u_{M+1} u_{M+2} \cdots u_N$.
$\times_{\hbar}$ is defined by
\begin{eqnarray}
&&(u_1 u_2 \cdots u_M) \times_{\hbar} (u_{M+1}u_{M+2} \cdots u_{N}) \n
&:=&T(u_1 \otimes u_2 \otimes \cdots \otimes u_{M} \otimes u_{M+1} \otimes \cdots \otimes u_{N}) \n
&:=&
\frac{1}{N!}\sum_{\sigma \in S_N} u_{\sigma_1} * u_{\sigma_2}*  \cdots * u_{\sigma_N}, \label{quantumproduct3}
\end{eqnarray}
where $u_1 \otimes u_2 \otimes \cdots  \otimes u_{N}$ is the symmetric tensor product. $S_N$ is the permutation group of $\{1,2, \cdots, N\}$. $*$ is the Moyal product defined by
\begin{equation}
f*g=\sum_{r=0}^{\infty} \frac{\alpha^r}{r!} 
\epsilon^{i_1 j_1} \epsilon^{i_2 j_2} \cdots \epsilon^{i_r j_r}
\frac{\partial}{\partial x_{i_1}} \frac{\partial}{\partial x_{i_2}} \cdots \frac{\partial}{\partial x_{i_r}} f
\frac{\partial}{\partial x_{j_1}} \frac{\partial}{\partial x_{j_2}} \cdots \frac{\partial}{\partial x_{j_r}}g,
\end{equation}
where $i_r$ and $j_r$ run from 1 to 2.
$\zeta$ is defined by 
\begin{equation}
\zeta(\sum_{r=0}^{\infty} \alpha^r u_r)
=
\sum_{r=0}^{\infty} \alpha^r Z_{u_r}.
\end{equation}
The quantum Zariski product is also Abelian, distributive and associative as follows. It is Abelian because
\begin{eqnarray}
Z_u \bullet_{\hbar} Z_v
&=&
ab \zeta((u_1 u_2 \cdots u_M) \times_{\hbar} (u_{M+1}u_{M+2} \cdots u_{N})) \n
&=&
ba \zeta( (u_{M+1}u_{M+2} \cdots u_{N}) \times_{\hbar} (u_1 u_2 \cdots u_M) ) \n
&=&
Z_v \bullet_{\hbar} Z_u,
\end{eqnarray}
and
\begin{eqnarray}
\bold{X}_{\hbar} \bullet_{\hbar} \bold{X}_{\hbar}'
&=& \sum_{u_0, v_0} Y^0_{u_0}(\sigma) Y'^0_{v_0}(\sigma)  Z_{u_0} \bullet_{\hbar} Z_{v_0} \n
&=& \sum_{v_0, u_0} Y'^0_{v_0}(\sigma) Y^0_{u_0}(\sigma) Z_{v_0} \bullet_{\hbar} Z_{u_0} \n
&=&\bold{X}_{\hbar}' \bullet_{\hbar} \bold{X}_{\hbar}.
\end{eqnarray}
It is distributive because
\begin{eqnarray}
\bold{X}_{\hbar} \bullet_{\hbar} (\bold{X}_{\hbar}'+\bold{X}_{\hbar}'') &=& (\sum_{r=0}^{\infty} \alpha^r \sum_{u_r} Y^r_{u_r}(\sigma) Z_{u_r})
\bullet_{\hbar} 
(\sum_{s=0}^{\infty} \alpha^s \sum_{v_s} Y'^s_{v_s}(\sigma) Z_{v_s}
+
\sum_{s=0}^{\infty} \alpha^s \sum_{w_s} Y''^s_{w_s}(\sigma) Z_{w_s}) \n
&=& (\sum_{u_0} Y^0_{u_0}(\sigma) Z_{u_0})
\bullet_{\hbar} 
(\sum_{v_0} Y'^0_{v_0}(\sigma) Z_{v_0}
+
\sum_{w_0} Y''^0_{w_0}(\sigma) Z_{w_0}) \n
&=& \sum_{u_0, v_0} Y^0_{u_0}(\sigma) Y'^0_{v_0}(\sigma)  Z_{u_0} \bullet_{\hbar} Z_{v_0}
+
\sum_{u_0, w_0} Y^0_{u_0}(\sigma) Y''^0_{w_0}(\sigma)  Z_{u_0} \bullet_{\hbar} Z_{w_0} \n
&=&
\bold{X}_{\hbar} \bullet_{\hbar} \bold{X}_{\hbar}'+ \bold{X}_{\hbar} \bullet_{\hbar} \bold{X}_{\hbar}''.
\end{eqnarray}
Associativity is verified as
\begin{eqnarray}
(Z_u \bullet_{\hbar} Z_v) \bullet_{\hbar} Z_w
&=&
ab \zeta((u_1 u_2 \cdots u_M) \times_{\hbar} (u_{M+1}u_{M+2} \cdots u_{N}))|_{\alpha=0} \bullet_{\hbar} Z_w \n
&=&
Z_{uv} \bullet_{\hbar} Z_w \n
&=&
abc \zeta((u_1 u_2 \cdots u_N) \times_{\hbar} (u_{N+1}u_{N+2} \cdots u_{L}))\n
&=&
abc \zeta(T(u_1 \otimes \cdots \otimes u_L)) \n
&=&
abc \zeta((u_1 u_2 \cdots u_M) \times_{\hbar} (u_{M+1}u_{M+2} \cdots u_{L}))\n
&=&
Z_{u} \bullet_{\hbar} Z_{vw} \n
&=&
Z_u \bullet_{\hbar} (Z_v \bullet_{\hbar} Z_w),
\end{eqnarray}
where $w=c u_{N+1} u_{N+2} \cdots u_{L}$, and thus
\begin{eqnarray}
(\bold{X}_{\hbar} \bullet_{\hbar} \bold{X}_{\hbar}') \bullet_{\hbar} \bold{X}_{\hbar}''
&=& \sum_{u_0, v_0, w_0} Y^0_{u_0}(\sigma) Y'^0_{v_0}(\sigma) Y''^0_{w_0}(\sigma)  (Z_{u_0} \bullet_{\hbar} Z_{v_0}) \bullet_{\hbar} Z_{w_0} \n
&=& \sum_{u_0, v_0, w_0} Y^0_{u_0}(\sigma) Y'^0_{v_0}(\sigma) Y''^0_{w_0}(\sigma)  Z_{u_0} \bullet_{\hbar} (Z_{v_0} \bullet_{\hbar} Z_{w_0}) \n
&=&\bold{X}_{\hbar} \bullet_{\hbar} (\bold{X}_{\hbar}' \bullet_{\hbar} \bold{X}_{\hbar}'').
\end{eqnarray}
Derivatives are defined as in the previous section by
\begin{equation}
\frac{\partial}{\partial \sigma^i} \bold{X}_{\hbar}=\sum_{r=0}^{\infty} \alpha^r \sum_{u_r} \frac{\partial}{\partial \sigma^i} Y^r_{u_r}(\sigma) Z_{u_r}.
\end{equation}
These derivatives are also commutative:
\begin{eqnarray}
\frac{\partial}{\partial \sigma^i} \frac{\partial}{\partial \sigma^j} \bold{X}_{\hbar}
&=&
\sum_{r=0}^{\infty} \alpha^r \sum_{u_r} \frac{\partial}{\partial \sigma^i} \frac{\partial}{\partial \sigma^j} Y^r_{u_r}(\sigma) Z_{u_r} \n
&=&
\sum_{r=0}^{\infty} \alpha^r \sum_{u_r} \frac{\partial}{\partial \sigma^j} \frac{\partial}{\partial \sigma^i} Y^r_{u_r}(\sigma) Z_{u_r} \n
&=&
\frac{\partial}{\partial \sigma^j} \frac{\partial}{\partial \sigma^i} \bold{X}_{\hbar},
\end{eqnarray}
and the derivatives of the quantum Zariski product also satisfy the Leibniz rule:
\begin{eqnarray}
\frac{\partial}{\partial \sigma^i} (\bold{X}_{\hbar} \bullet_{\hbar} \bold{X}_{\hbar}')
&=&
\sum_{u_0, v_0} \frac{\partial}{\partial \sigma^i} (Y^0_{u_0}(\sigma) Y'^0_{v_0}(\sigma))  Z_{u_0} \bullet_{\hbar} Z_{v_0} \n
&=&
\sum_{u_0, v_0}( \frac{\partial}{\partial \sigma^i} Y^0_{u_0}(\sigma) Y'^0_{v_0}(\sigma) + Y^0_{u_0}(\sigma) \frac{\partial}{\partial \sigma^i} Y'^0_{v_0}(\sigma))  Z_{u_0} \bullet_{\hbar} Z_{v_0} \n
&=&
\frac{\partial}{\partial \sigma^i} \bold{X}_{\hbar} \bullet_{\hbar} \bold{X}_{\hbar}'
+ \bold{X}_{\hbar} \bullet_{\hbar} 
\frac{\partial}{\partial \sigma^i} \bold{X}_{\hbar}'.
\end{eqnarray}

We define the Zariski quantized Nambu-Poisson bracket by
\begin{eqnarray}
[\bold{X}_{\hbar}, \bold{X}_{\hbar}', \bold{X}_{\hbar}'']_{\bullet_{\hbar}}
&:=&\epsilon^{ijk}
\frac{\partial}{\partial\sigma^i} \bold{X}_{\hbar} \bullet_{\hbar}
\frac{\partial}{\partial\sigma^j} \bold{X}_{\hbar}' \bullet_{\hbar}
\frac{\partial}{\partial\sigma^k} \bold{X}_{\hbar}'' \n
&=&
\sum_{u_0,v_0,w_0}
\epsilon^{ijk}
\frac{\partial}{\partial\sigma^i} Y^0_{u_0}(\sigma)
\frac{\partial}{\partial\sigma^j} Y'^0_{v_0}(\sigma)
\frac{\partial}{\partial\sigma^k} Y''^0_{w_0}(\sigma)
Z_{u_0} \bullet_{\hbar} Z_{v_0} \bullet_{\hbar} Z_{w_0},
\end{eqnarray}
where $i, j, k= 1, 2, 3$.
By definition, the bracket is skew-symmetric. One can show that it satisfies the Fundamental Identity and the Leibniz rule as exactly in the same way as in the previous section by using the above properties. Thus, the Zariski quantized Nambu-Poisson bracket, which is a deformation quantization of the many-body deformation of the original Nambu-Poisson bracket, has the same Nambu-Poisson structure as the original Nambu-Poison bracket.

We can show that a quantum Zariski product of two elements depends only on $\alpha^2$ in the following way.
Because 
\begin{eqnarray}
&&v_{\sigma^1} * v_{\sigma^2} * \cdots * v_{\sigma^N} \n
&=&
\sum_{n_1, \cdots, n_{N-1}} \alpha^{n_1+ \cdots + n_{N-1}}\frac{1}{n_1! \cdots n_{N-1}!}(\epsilon^{i_{1, 1}j_{1, 1}} \cdots \epsilon^{i_{1, n_1} j_{1, n_1}})  
(\epsilon^{i_{2, 1}j_{2, 1}} \cdots \epsilon^{i_{2, n_2} j_{2, n_2}}) \cdots
\epsilon^{i_{N-1, n_{N-1}} j_{N-1, n_{N-1}}}  \n
&&\partial_{i_{N-1, 1}} \cdots \partial_{i_{N-1, n_{N-1}}}( \cdots (\partial_{i_{2, 1}} \cdots \partial_{i_{2, n_2}}(\partial_{i_{1, 1}} \cdots \partial_{i_{1, n_1}} v_{\sigma^1}
\partial_{j_{1, 1}} \cdots \partial_{j_{1, n_1}} v_{\sigma^2})
\partial_{j_{2, 1}} \cdots \partial_{j_{2, n_2}} v_{\sigma^3}) \n
&& \cdots )\partial_{j_{N-1, 1}} \cdots \partial_{j_{N-1, n_{N-1}}} v_{\sigma^N} \n
&=&
\sum_{n_1, \cdots, n_{N-1}} (-\alpha)^{n_1+ \cdots + n_{N-1}}\frac{1}{n_1! \cdots n_{N-1}!}(\epsilon^{j_{N-1, 1}i_{N-1, 1}} \cdots \epsilon^{j_{N-1, n_{N-1}} i_{N-1, n_{N-1}}}) \cdots  
(\epsilon^{j_{1, 1}i_{1, 1}} \cdots \epsilon^{j_{1, n_1} i_{1, n_1}})   \n
&&\partial_{j_{N-1, 1}} \cdots \partial_{j_{N-1, n_{N-1}}}v_{\sigma^N} \partial_{i_{N-1, 1}} \cdots \partial_{i_{N-1, n_{N-1}}}(\partial_{j_{N-2, 1}} \cdots \partial_{j_{N-2, n_{N-2}}} v_{\sigma^{N-1}}
\partial_{i_{N-2, 1}} \cdots \partial_{i_{N-2, n_{N-2}}}( \n
&&
\partial_{j_{N-3, 1}} \cdots \partial_{j_{N-3, n_{N-3}}} v_{\sigma^{N-2}}
\partial_{i_{N-3, 1}} \cdots \partial_{i_{N-3, n_{N-3}}}( \cdots  
(\partial_{j_{1, 1}} \cdots \partial_{j_{1, n_1}}
 v_{\sigma^2} 
\partial_{i_{1, 1}} \cdots \partial_{i_{1, n_1}}
 v_{\sigma^1})) \cdots ), \n
\end{eqnarray}
and
\begin{eqnarray}
&&v_{\sigma^N} * v_{\sigma^{N-1}} * \cdots * v_{\sigma^1} \n
&=&
\sum_{n_1, \cdots, n_{N-1}} (\alpha)^{n_1+ \cdots + n_{N-1}}\frac{1}{n_1! \cdots n_{N-1}!}(\epsilon^{j_{N-1, 1}i_{N-1, 1}} \cdots \epsilon^{j_{N-1, n_{N-1}} i_{N-1, n_{N-1}}}) \cdots  
(\epsilon^{j_{1, 1}i_{1, 1}} \cdots \epsilon^{j_{1, n_1} i_{1, n_1}})   \n
&&\partial_{j_{N-1, 1}} \cdots \partial_{j_{N-1, n_{N-1}}}v_{\sigma^N} \partial_{i_{N-1, 1}} \cdots \partial_{i_{N-1, n_{N-1}}}(\partial_{j_{N-2, 1}} \cdots \partial_{j_{N-2, n_{N-2}}} v_{\sigma^{N-1}}
\partial_{i_{N-2, 1}} \cdots \partial_{i_{N-2, n_{N-2}}}( \n
&&
\partial_{j_{N-3, 1}} \cdots \partial_{j_{N-3, n_{N-3}}} v_{\sigma^{N-2}}
\partial_{i_{N-3, 1}} \cdots \partial_{i_{N-3, n_{N-3}}}( \cdots  
(\partial_{j_{1, 1}} \cdots \partial_{j_{1, n_1}}
 v_{\sigma^2} 
\partial_{i_{1, 1}} \cdots \partial_{i_{1, n_1}}
 v_{\sigma^1})) \cdots ), \n
\end{eqnarray}
we obtain 
\begin{eqnarray}
&&\frac{1}{N!} \sum_{\sigma \in S_N} (v_{\sigma^1} * \cdots *v_{\sigma^N}) \n
&=& \frac{1}{2} \frac{1}{N!} \sum_{\sigma \in S_N} (v_{\sigma^1} * \cdots *v_{\sigma^N}+
v_{\sigma^N} * \cdots *v_{\sigma^1}) \n
&=& \frac{1}{N!} \sum_{\sigma \in S_N}
\sum_{n_1+ \cdots +n_{N-1}=2n} (\alpha)^{2n}\frac{1}{n_1! \cdots n_{N-1}!}(\epsilon^{j_{N-1, 1}i_{N-1, 1}} \cdots \epsilon^{j_{N-1, n_{N-1}} i_{N-1, n_{N-1}}}) \cdots  
(\epsilon^{j_{1, 1}i_{1, 1}} \cdots \epsilon^{j_{1, n_1} i_{1, n_1}})   \n
&&\partial_{j_{N-1, 1}} \cdots \partial_{j_{N-1, n_{N-1}}}v_{\sigma^N} \partial_{i_{N-1, 1}} \cdots \partial_{i_{N-1, n_{N-1}}}(\partial_{j_{N-2, 1}} \cdots \partial_{j_{N-2, n_{N-2}}} v_{\sigma^{N-1}}
\partial_{i_{N-2, 1}} \cdots \partial_{i_{N-2, n_{N-2}}}( \n
&&
\partial_{j_{N-3, 1}} \cdots \partial_{j_{N-3, n_{N-3}}} v_{\sigma^{N-2}}
\partial_{i_{N-3, 1}} \cdots \partial_{i_{N-3, n_{N-3}}}( \cdots  
(\partial_{j_{1, 1}} \cdots \partial_{j_{1, n_1}}
 v_{\sigma^2} 
\partial_{i_{1, 1}} \cdots \partial_{i_{1, n_1}}
 v_{\sigma^1})) \cdots ). \n \label{alpha=hbar^2}
\end{eqnarray}
From (\ref{quantumproduct1}), (\ref{quantumproduct2}), (\ref{quantumproduct3}), and (\ref{alpha=hbar^2}), the statement is proven true. Therefore, we identify $\alpha^2$ as $\hbar$.

\subsection{Metric}
In this subsection, we construct a natural metric for $\mathcal{M}_{\hbar}$. In particular, the metric is invariant under a gauge transformation generated by the Zariski quantized Nambu-Poisson bracket.

We define a metric for $X_{\hbar}, X_{\hbar}' \in \mathcal{M}_{\hbar}$ by
\begin{eqnarray}
<\bold{X}_{\hbar}, \bold{X}_{\hbar}'> 
&=&
<\bold{X}_{\hbar} \bullet_{\hbar}  \bold{X}_{\hbar}'> \n
&=&
\int d^p\sigma <<\bold{X}_{\hbar} \bullet_{\hbar} \bold{X}_{\hbar}'>> \n
&=&
\sum_{u_0, v_0} \int d^p \sigma Y^0_{u_0}(\sigma) Y'^0_{v_0}(\sigma) <<Z_{u_0} \bullet_{\hbar} Z_{v_0}>> \n
&=&
\sum_{u_0, v_0} \int d^p \sigma Y^0_{u_0}(\sigma) Y'^0_{v_0}(\sigma) \sum_{r=0}^{\infty} \alpha^r \sum_{w_r} <<Z_{w_r}>>,
\end{eqnarray}
where $<<Z_{w_r}>>$ are defined in the same way as in the subsection 2.2.
This metric is invariant under a gauge transformation generated by the $p$-dimensional Zariski quantized Nambu-Poisson bracket. Here we show the $p=3$ case as an example, whereas the $p=2$ and $p>3$ case are shown in a similar way. The condition of the invariance of the metric is given by 
\begin{equation}
<[\bold{X}_{\hbar}^3, \bold{X}_{\hbar}^4, \bold{X}_{\hbar}^1]_{\bullet_{\hbar}}, \bold{X}_{\hbar}^2> + <\bold{X}_{\hbar}^1, [\bold{X}_{\hbar}^3, \bold{X}_{\hbar}^4, \bold{X}_{\hbar}^2]_{\bullet_{\hbar}}>=0,
\end{equation}
in the same way as in the previous case.
The left hand side is
\begin{eqnarray}
&&<\sum_{(u_3)_0,(u_4)_0,(u_1)_0}
\epsilon^{ijk}
\partial_i (Y^3)^0_{(u_{3})_0}
\partial_j (Y^4)^0_{(u_4)_0}
\partial_k (Y^1)^0_{(u_1)_0}
Z_{(u_3)_0} \bullet_{\hbar} Z_{(u_4)_0} \bullet_{\hbar} Z_{(u_1)_0},
\sum_{(u_2)_0} (Y^2)^0_{(u_2)_0} Z_{(u_2)_0}> \n
&+&
<\sum_{(u_1)_0} (Y^1)^0_{(u_1)_0}  Z_{(u_1)_0},
\sum_{(u_3)_0,(u_4)_0,(u_2)_0}
\epsilon^{ijk}
\partial_i (Y^3)^0_{(u_3)_0}
\partial_j (Y^4)^0_{(u_4)_0}
\partial_k (Y^2)^0_{(u_2)_0}
Z_{(u_3)_0} \bullet_{\hbar} Z_{(u_4)_0} \bullet_{\hbar} Z_{(u_2)_0}> \n
&=&
\sum_{(u_1)_0, (u_2)_0, (u_3)_0, (u_4)_0}
\int d^3 \sigma (\epsilon^{ijk}
\partial_i (Y^3)^0_{(u_3)_0}
\partial_j (Y^4)^0_{(u_4)_0}
\partial_k (Y^1)^0_{(u_1)_0}
(Y^2)^0_{(u_2)_0} \n
&& \qquad \qquad +
(Y^1)^0_{(u_1)_0} 
\epsilon^{ijk}
\partial_i (Y^3)^0_{(u_3)_0}
\partial_j (Y^4)^0_{(u_4)_0}
\partial_k (Y^2)^0_{(u_2)_0})
<<Z_{(u_1)_0} \bullet_{\hbar} Z_{(u_2)_0} \bullet_{\hbar} Z_{(u_3)_0} \bullet_{\hbar} Z_{(u_4)_0}>> \n
&=&
\sum_{(u_1)_0, (u_2)_0, (u_3)_0, (u_4)_0}
\int d^3 \sigma (
\partial_k
\epsilon^{ijk}
\partial_i (Y^3)^0_{(u_3)_0}
\partial_j (Y^4)^0_{(u_4)_0}
(Y^1)^0_{(u_1)_0}
(Y^2)^0_{(u_2)_0}) \n
&& \qquad \qquad \qquad \qquad \qquad \quad <<Z_{(u_1)_0} \bullet_{\hbar} Z_{(u_2)_0} \bullet_{\hbar} Z_{(u_3)_0} \bullet_{\hbar} Z_{(u_4)_0}>> \n
&=&0.
\end{eqnarray}

\subsection{Pair Creation and Annihilation}
In this subsection, we study general features of Zariski quantized theories by using a simple model. We start with a first quantized action,
\begin{equation}
S_0=\frac{1}{2} X^2
+ \lambda X^6, \label{firstquantizedaction}
\end{equation}
where $X \in \bold{R}$ represents a target coordinate.  
If we deform it by the classical Zariski product and $\mathcal{M}$, we obtain 
\begin{equation}
S=<\frac{1}{2} \bold{X} \bullet \bold{X}
+ \lambda (\bold{X})_{\bullet}^6>, \label{classicalaction}
\end{equation}
where $\bold{X}=\sum_u Y_u Z_u$ and 
$Y_u \in \bold{R}$.
After the Zariski quantization, we obtain
\begin{equation}
S_{\hbar}=<\frac{1}{2} \bold{X} \bullet_{\hbar} \bold{X} + \lambda (\bold{X})^6_{\bullet_{\hbar}}>.  \label{Zariskiquantizedaction}
\end{equation}
We define a theory by a path-integral as
\begin{equation}
Z= \int \mathcal{D}Y \exp{(\frac{i}{\hbar}<\frac{1}{2} \bold{X} \bullet_{\hbar} \bold{X}
+ \lambda (\bold{X})_{\bullet_{\hbar}}^6>)}. \label{pathint}
\end{equation}
(\ref{classicalaction}) is a classical action of this theory because (\ref{classicalaction}) dominates in (\ref{pathint}) in the $\hbar \to 0$ limit. 

\begin{figure}
\begin{center}
\subfigure[a classical interaction]{
\psfrag{x1}{$Y_{x_1} Z_{x_1}$}
\psfrag{x2}{$Y_{x_1} Z_{x_1}$}
\psfrag{x3}{$Y_{x_2} Z_{x_2}$}
\psfrag{x4}{$Y_{x_2} Z_{x_2}$}
\psfrag{x5}{$Y_{x_2^2} Z_{x_2^2}$}
\psfrag{x6}{$Y_{x_1^2} Z_{x_1^2}$}
\includegraphics[height=5cm, keepaspectratio, clip]{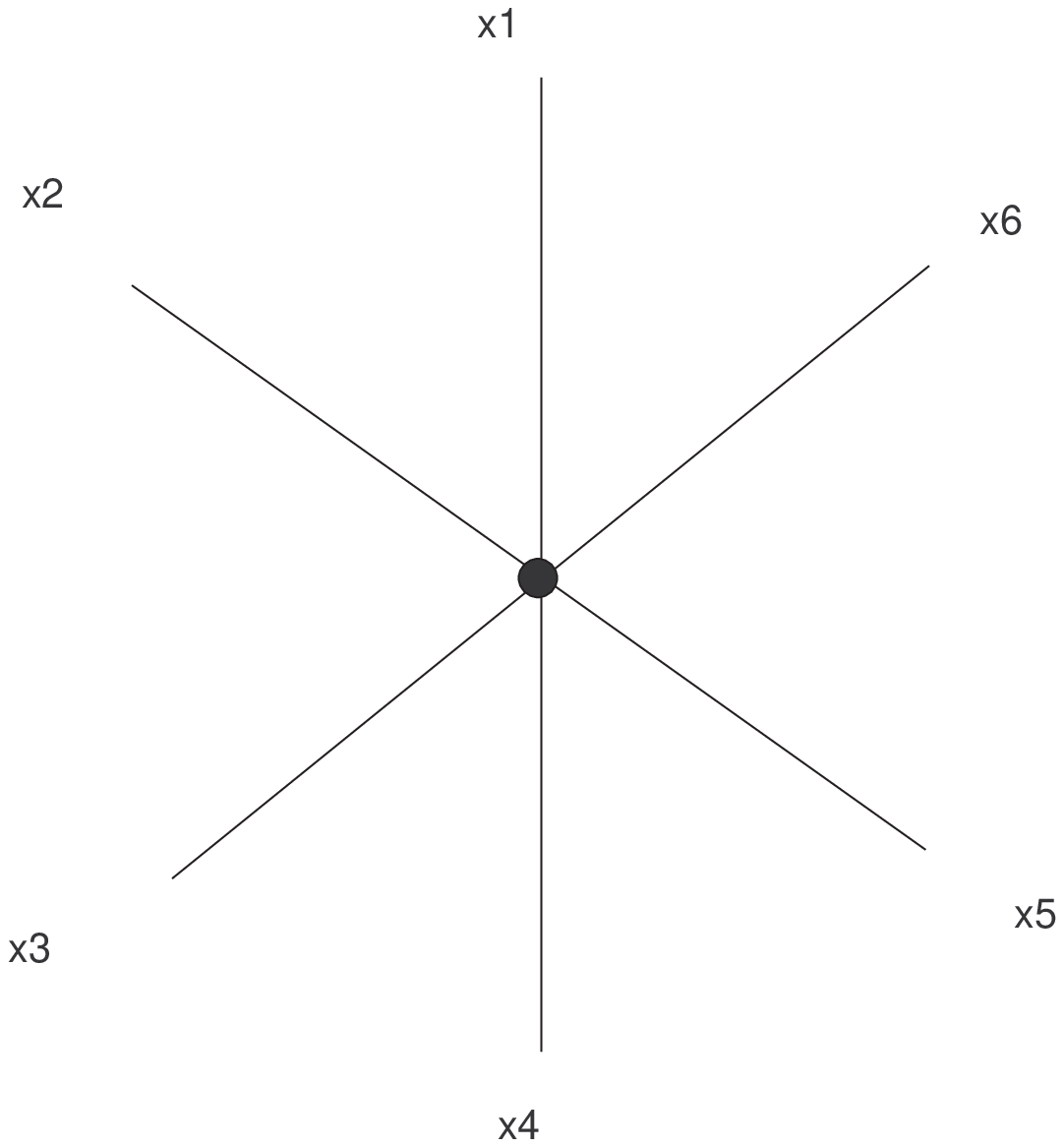}
\label{classical}
}
\hfill
\subfigure[a quantum correction]{
\psfrag{x1}{$Y_{x_1} Z_{x_1}$}
\psfrag{x2}{$Y_{x_1} Z_{x_1}$}
\psfrag{x3}{$Y_{x_2} Z_{x_2}$}
\psfrag{x4}{$Y_{x_2} Z_{x_2}$}
\psfrag{x5}{$Y_{x_1^2}Y_{x_2^2}$}
\psfrag{x6}{$\hbar$}
\includegraphics[height=5cm, keepaspectratio, clip]{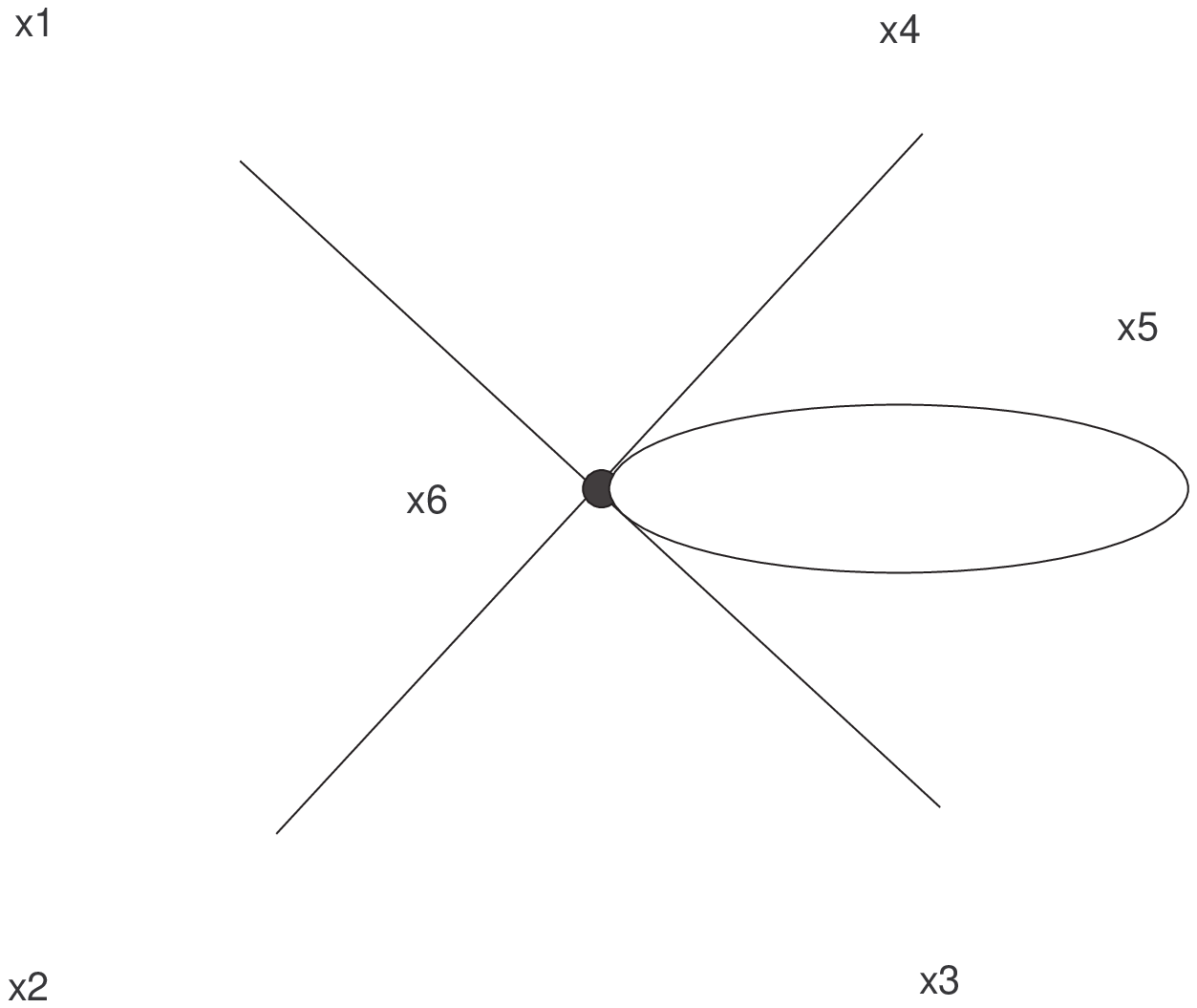}
\label{quantum}
}
\end{center}
\caption{typical interactions}
\end{figure}

We have typical interaction terms in (\ref{Zariskiquantizedaction});
\begin{eqnarray}
&&\lambda (Y_{x_1} Z_{x_1}) \bullet_{\hbar} (Y_{x_1} Z_{x_1}) \bullet_{\hbar} (Y_{x_1^2} Z_{x_1^2}) \bullet_{\hbar} (Y_{x_2} Z_{x_2}) \bullet_{\hbar} (Y_{x_2} Z_{x_2}) \bullet_{\hbar} (Y_{x_2^2} Z_{x_2^2}) \n
&=& \lambda Y_{x_1} Y_{x_1} Y_{x_1^2} Y_{x_2} Y_{x_2} Y_{x_2^2} Z_{x_1} \bullet_{\hbar} Z_{x_1} \bullet_{\hbar} Z_{x_1^2} \bullet_{\hbar} Z_{x_2} \bullet_{\hbar} Z_{x_2} \bullet_{\hbar} Z_{x_2^2} \n
&=& \lambda Y_{x_1} Y_{x_1} Y_{x_1^2} Y_{x_2} Y_{x_2} Y_{x_2^2} \zeta(T(x_1 \otimes x_1 \otimes x_1 \otimes x_1 \otimes x_2  \otimes x_2 \otimes x_2 \otimes x_2)) \n
&=& \lambda Y_{x_1} Y_{x_1} Y_{x_1^2} Y_{x_2} Y_{x_2} Y_{x_2^2} 
\zeta(x_1^4 x_2^4 + \frac{16}{5} \hbar x_1^2x_2^2 + O(\hbar^2) ) \n
&=& \lambda Y_{x_1} Y_{x_1} Y_{x_1^2} Y_{x_2} Y_{x_2} Y_{x_2^2} 
(Z_{x_1}Z_{x_1}Z_{x_1^2}Z_{x_2}Z_{x_2}Z_{x_2^2}
+ \frac{16}{5} \hbar Z_{x_1}Z_{x_1}Z_{x_2}Z_{x_2}+O(\hbar^2)) \n
&=& \lambda (Y_{x_1} Z_{x_1}) (Y_{x_1} Z_{x_1}) (Y_{x_1^2} Z_{x_1^2}) (Y_{x_2} Z_{x_2}) (Y_{x_2} Z_{x_2}) (Y_{x_2^2} Z_{x_2^2}) \n
&&+ \frac{16}{5} \hbar \lambda (Y_{x_1^2}Y_{x_2^2}) (Y_{x_1} Z_{x_1}) (Y_{x_1} Z_{x_1}) (Y_{x_2} Z_{x_2}) (Y_{x_2} Z_{x_2}) +O(\hbar^2). \label{interactions}
\end{eqnarray}
In (\ref{interactions}), $\lambda (Y_{x_1} Z_{x_1}) (Y_{x_1} Z_{x_1}) (Y_{x_1^2} Z_{x_1^2}) (Y_{x_2} Z_{x_2}) (Y_{x_2} Z_{x_2}) (Y_{x_2^2} Z_{x_2^2})$ is a classical interaction term, whereas  $\frac{16}{5} \hbar \lambda (Y_{x_1^2}Y_{x_2^2}) (Y_{x_1} Z_{x_1}) (Y_{x_1} Z_{x_1}) (Y_{x_2} Z_{x_2}) (Y_{x_2} Z_{x_2})$ is a quantum correction. Because each $Z_u$ in the interactions is a base for each one-body state, classical interactions represent 6-body interactions as shown as an example in Fig.\ref{classical}, whereas the quantum correction represents a 4-body interaction.  Because the quantum correction is $O(\hbar)$, it should be interpreted to be an one-loop correction as shown in Fig.\ref{quantum}. This implies that pair creations and annihilations occur among the many bodies introduced by the many-body deformation. Therefore, we conclude that first quantized field theories become second quantized theories after the Zariski quantization. 

\section{Examples}
The Zariski quantization is applicable to any first quantized field theories and preserves the supersymmetries of them. In this section, we present relevant examples.

The Zariski quantized type IIB superstring action \cite{GreenSchwarz} is given by
\begin{eqnarray}
S_{s}&=&-\frac{1}{2 \pi \alpha'} < \sqrt{-\frac{1}{2}(\epsilon^{ij}\bold{\Pi}^{\mu}_i \bullet_{\hbar} \bold{\Pi}^{\nu}_j)^2_{\bullet_{\hbar}}}_{\bullet_{\hbar}}
+i \epsilon^{ij} \partial_i \bold{X}^{\mu} \bullet_{\hbar} (\bar{\bold{\Theta}}^1 \bullet_{\hbar} \Gamma_{\mu} \partial_j \bold{\Theta}^1 - \bar{\bold{\Theta}}^2 \bullet_{\hbar} \Gamma_{\mu} \partial_j \bold{\Theta}^2) \n
&&-\epsilon^{ij}\bar{\bold{\Theta}}^1 \bullet_{\hbar} \Gamma^{\mu} \partial_i \bold{\Theta}^1 \bullet_{\hbar} \bar{\bold{\Theta}}^2 \bullet_{\hbar} \Gamma_{\mu} \partial_j \bold{\Theta}^2>,
\end{eqnarray} 
where 
$i, j = 0, 1$, $\mu, \nu =0, \cdots, 9$, 
$\bold{\Pi}^{\mu}_i = \partial_i \bold{X}^{\mu} -i \bar{\bold{\Theta}}^1 \bullet_{\hbar} \Gamma^{\mu} \partial_i \bold{\Theta}^1 -
i \bar{\bold{\Theta}}^2 \bullet_{\hbar} \Gamma^{\mu} \partial_i \bold{\Theta}^2$. $\bold{\Theta}^1$ and $\bold{\Theta}^2$ are $SO(1, 9)$ Majorana-Weyl fermions that possess the same chirality.
The Zariski quantized supermembrane action \cite{BergshoeffSezginTownsend} is given by
\begin{eqnarray}
S_{M}&=&
\Bigl<
\sqrt{-det\bold{G}_{\bullet_{\hbar}}}_{\bullet_{\hbar}}
+\frac{i}{4} \epsilon^{i j k} \bar{\bold{\Psi}} \bullet_{\hbar} \Gamma_{MN} \partial_{i} \bold{\Psi} \bullet_{\hbar}
(\bold{\Pi}_{j}^{\,\,\, M} \bullet_{\hbar} \bold{\Pi}_{k}^{\,\,\, N}
+\frac{i}{2}\bold{\Pi}_{j}^{\,\,\, M} \bullet_{\hbar} \bar{\bold{\Psi}} \bullet_{\hbar} \Gamma^N \partial_{k}\bold{\Psi} \n
&& \qquad \qquad \qquad \qquad \qquad \qquad \qquad -\frac{1}{12} \bar{\bold{\Psi}} \bullet_{\hbar} \Gamma^M \partial_{j}\bold{\Psi} \bullet_{\hbar} \bar{\bold{\Psi}} \bullet_{\hbar} \Gamma^N \partial_{k}\bold{\Psi})
\Bigr>, 
\end{eqnarray}
where $i, j, k = 0, 1, 2$, $M, N=0, \cdots, 10$, $\bold{G}_{i j}= \bold{\Pi}_{i}^{\,\,\, M} {\bullet_{\hbar}} \bold{\Pi}_{j M}$ and $\bold{\Pi}_{i}^{\,\,\, M}= \partial_{i} \bold{X}^{M}
-\frac{i}{2} \bar{\bold{\Psi}} {\bullet_{\hbar}} \Gamma^{M} \partial_{i} \bold{\Psi}$.
$\bold{\Psi}$ is a $SO(1, 10)$ Majorana fermion.
These two theories are expected to be second quantized covariant theories of superstrings and supermembranes.

By performing the Zariski quantization of the type IIB superstring in the Schild gauge \cite{Schild}, which is equivalent to the IIB matrix model \cite{IKKT} with the area preserving diffeomorphism symmetry, we obtain
\begin{equation}
S_{IIB} = < -\frac{1}{4} \{ \bold{X}^{\mu}, \bold{X}^{\nu} \}_{\bullet_{\hbar}}^2
- \frac{1}{2} \bar{\bold{\Theta}} {\bullet_{\hbar}} \Gamma^{\mu} \{\bold{X}_{\mu}, \bold{\Theta} \}_{\bullet_{\hbar}}>, \label{IIBaction}
\end{equation}
where $\bold{\Theta}$ is a $SO(1, 9)$ Majorana-Weyl fermions.
By performing the Zariski quantization of the supermembrane action in a semi-light-cone gauge \cite{MModel}, which is equivalent to the 3-algebra model of M-theory with the volume preserving diffeomorphism symmetry \cite{MModel, LorentzianM}, we obtain
\begin{eqnarray}
S_{3algM}
&=&
\Bigl<
-\frac{1}{12}\{\bold{X}^I, \bold{X}^J, \bold{X}^K\}_{\bullet_{\hbar}}^2 
-\frac{1}{2}(\bold{A}^{u}_{\alpha a b} {\bullet_{\hbar}} \{\varphi^a_u, \varphi^b_u, \bold{X}^{I}\})_{\bullet_{\hbar}}^2  \n
&& \qquad \qquad
-\frac{1}{3} E^{\alpha \beta \gamma} 
\bold{A}^u_{\alpha a b} {\bullet_{\hbar}} \bold{A}^v_{\beta c d} {\bullet_{\hbar}} \bold{A}^w_{\gamma e f} 
\{\varphi^a_u, \varphi^c_v, \varphi^d_v\}\{\varphi^b_u, \varphi^e_w, \varphi^f_w\}  \n
&& \qquad \qquad
-\frac{i}{2}\bar{\bold{\Psi}} {\bullet_{\hbar}} \Gamma^{\alpha} \bold{A}^u_{\alpha a b} {\bullet_{\hbar}} \{\varphi^a_u, \varphi^b_u, \bold{\Psi}\}
+\frac{i}{4}\bar{\bold{\Psi}} {\bullet_{\hbar}} \Gamma_{IJ}\{\bold{X}^I, \bold{X}^J, \bold{\Psi}\}_{\bullet_{\hbar}} 
\Bigr>, 
\label{continuumaction}
\end{eqnarray}
where $\alpha, \beta, \gamma = 0, 1, 2$, $I, J, K = 3, \cdots, 10$ and $\varphi^a$ are complete basis of functions in three-dimensions. 
$E^{\alpha \beta \gamma}$ is a Levi-Civita symbol in three dimensions. 
$\bold{\Psi}$ is a $SO(1,2) \times SO(8)$ Majorana-Weyl fermion satisfying
\begin{eqnarray}
&&\Gamma^{012}\bold{\Psi}=-\bold{\Psi} \label{kappa}, \\
&& \bold{\Psi}^{\dagger} = \bold{\Psi}^{T}.
\end{eqnarray}
These theories are also expected to be second quantized theories of superstrings and supermembranes. It is rather easy to study their relations to matrix models and string field theories because (\ref{IIBaction}) and (\ref{continuumaction}) possess large gauge symmetries, the area and volume preserving diffeomorphism symmetry, respectively, which should correspond to the large gauge symmetries of the matrix models and string field theories.





\section{Conclusion and Discussion}
\setcounter{equation}{0}
In this paper, we clarified physical meaning of the Zariski quantization. We found that we can obtain second quantized theories by performing the Zariski quantization, which consists of the many-body deformation and deformation quantization, from first-quantized field theories, such as superstring and supermembrane theories. The Zariski quantization preserves the supersymmetries of the first quantized theories, because the quantum Zariski product is Abelian, associative and distributive, and admits commutative derivative satisfying the Leibniz rule. Therefore, by performing  the Zariski quantization of superstring and supermembrane theory, we can obtain second quantized theories of superstring and supermembrane.  

We discuss the origin of the difference between the physical consequences in the paper by G. Dito et al. and in our paper. In our paper, we deformed the first quantized field theories by $\mathcal{M}$, which depend on both $\sigma$-spaces and $x$-spaces. In $\mathcal{M}$, $u(x)$ are labels on many bodies and $Y_u(\sigma)$ are their fields. As a result, Zariski quantized theories are second quantized theories. On the other hand, G. Dito et al. studied the Zariski quantization on small subspaces $\mathcal{A}_0$, which depend essentially only on $x$-spaces since the basis $J(Z_u) \in \mathcal{A}_0$ satisfy $\partial_{\sigma}J(Z_u)= J(Z_{\partial_{x}u})$ \cite{DitoFlatoSternheimerTakhtajan}. Then, there is no degree of freedom of $Y_u(\sigma)$ in $\mathcal{A}_0$. Because one needs to define physical observables by $u$ themselves in $\mathcal{A}_0$, Zariski quantized theories cannot be second quantized theories and the Zariski quantization was interpreted as "sesqui-quantization," a halfway between first and second quantizations.

String field theories and several matrix models are known to describe many-body strings or membranes although they have not been proved to formulate non-perturbative string theory yet. We hope that non-perturbative dynamics of string theory will be derived from the Zariski quantized superstring and supermembrane theories. One reasonable way is to study the relations among the Zariski quantized theories, the string field theories and the matrix models.


%
%
%


\end{document}